\documentclass[aps,prx,twocolumn,showpacs,superscriptaddress,floatfix ,nofootinbib]{revtex4-2}
\usepackage{amsmath}
\usepackage{graphicx}
\usepackage{dcolumn}
\usepackage{bm}
\usepackage[colorlinks,citecolor=blue,linkcolor=blue,anchorcolor=blue,filecolor=blue,
urlcolor=blue]{hyperref}
\usepackage{hyperref}
\usepackage{ulem}
\usepackage{xcolor}
\usepackage[utf8]{inputenc}
\usepackage{tabularx}
\usepackage{color}
\usepackage{cellspace}
\usepackage{lipsum, babel}
\usepackage{lipsum}
\usepackage{multirow}
\usepackage{booktabs} 
\definecolor{customviolet}{RGB}{148,0,211}
\usepackage{orcidlink}
\usepackage{float}
\usepackage{adjustbox}
\newcommand{\be}{\begin{equation}}
\newcommand{\ee}{\end{equation}}
\newcommand{\bea}{\begin{eqnarray}}
\newcommand{\eea}{\end{eqnarray}}

\usepackage{orcidlink}

\begin{document}

\preprint{APS/123-QED}


\title{Supernova Remnants with Mirror Dark Matter and Hyperons}

\author{Adamu Issifu~\orcidlink{0000-0002-2843-835X}} 
\email{ai@academico.ufpb.br}
\affiliation{Departamento de F\'isica e Laborat\'orio de Computa\c c\~ao Cient\'ifica Avan\c cada e Modelamento (Lab-CCAM), Instituto Tecnol\'ogico de Aeron\'autica, DCTA, 12228-900, S\~ao Jos\'e dos Campos, SP, Brazil} 

\author{Prashant Thakur~\orcidlink{0000-0003-4189-6176}}
\affiliation{Department of Physics, BITS-Pilani, K. K. Birla Goa Campus, Goa 403726, India}

\author{Franciele M. da Silva~\orcidlink{0000-0003-2568-2901}} 
\affiliation{Departamento de F\'isica, CFM - Universidade Federal de Santa Catarina; \\ C.P. 476, CEP 88.040-900, Florian\'opolis, SC, Brazil.}

\author{Kau D. Marquez~\orcidlink{0000-0001-5930-7179}}
\affiliation{Departamento de F\'isica e Laborat\'orio de Computa\c c\~ao Cient\'ifica Avan\c cada e Modelamento (Lab-CCAM), Instituto Tecnol\'ogico de Aeron\'autica, DCTA, 12228-900, S\~ao Jos\'e dos Campos, SP, Brazil (in loving memory)} 

\author{D\'ebora P. Menezes~\orcidlink{0000-0003-0730-6689}}
\affiliation{Departamento de F\'isica, CFM - Universidade Federal de Santa Catarina; \\ C.P. 476, CEP 88.040-900, Florian\'opolis, SC, Brazil.}

\author{M. Dutra~\orcidlink{0000-0001-7501-0404}} 
\affiliation{Departamento de F\'isica e Laborat\'orio de Computa\c c\~ao Cient\'ifica Avan\c cada e Modelamento (Lab-CCAM), Instituto Tecnol\'ogico de Aeron\'autica, DCTA, 12228-900, S\~ao Jos\'e dos Campos, SP, Brazil} 

\author{O. Louren\c{c}o~\orcidlink{0000-0002-0935-8565}}  
\affiliation{Departamento de F\'isica e Laborat\'orio de Computa\c c\~ao Cient\'ifica Avan\c cada e Modelamento (Lab-CCAM), Instituto Tecnol\'ogico de Aeron\'autica, DCTA, 12228-900, S\~ao Jos\'e dos Campos, SP, Brazil} 

\author{Tobias Frederico~\orcidlink{0000-0002-5497-5490}} 
\affiliation{Departamento de F\'isica e Laborat\'orio de Computa\c c\~ao Cient\'ifica Avan\c cada e Modelamento (Lab-CCAM), Instituto Tecnol\'ogico de Aeron\'autica, DCTA, 12228-900, S\~ao Jos\'e dos Campos, SP, Brazil}

\date{\today}
\begin{abstract}
For the first time, we use relativistic mean-field (RMF) approximation with density-dependent couplings, adjusted by the DDME2 parameterization, to investigate the effects of dark matter on supernova remnants. We calculate the nuclear equation of state for nuclear and dark matter separately, under the thermodynamic conditions related to the evolution of supernova remnants. A mirrored model is adopted for dark matter, and its effect on remnant matter is studied using a two-fluid scenario. At each stage of the remnant evolution, we assume that dark and ordinary matter have the same entropy and lepton fraction, and a fixed proportion of dark matter mass fraction is added to the stellar matter to observe its effects on some microscopic and macroscopic properties of the star. We observe that dark matter in the remnant core reduces the remnant's maximum mass, radius, and tidal deformability. Moreover, dark matter heats the remnant matter and alters particle distributions, thereby decreasing its isospin asymmetry and increasing the sound speed through the matter.

\end{abstract}

\maketitle
\section{Introduction}
The universe is thought to consist of approximately 6\% visible matter, while the remaining ~94\% is composed of dark matter (DM) and dark energy, consisting of 26\% and ~68\% of the total mass-energy content respectively~\cite{ParticleDataGroup:2022pth}. Compelling evidence for the existence of DM arises from observations such as galactic rotation curves, galaxy clusters, large-scale cosmological structures, and gravitational lensing. However, the precise nature of DM -- particularly its mass and interaction properties -- remains an open question under active investigation. Consequently, constraining DM properties through both direct and indirect approaches is a critical focus of contemporary research (see reviews in~\cite{Klasen:2015uma, Misiaszek:2023sxe}).

Given that the exact nature of DM and its in-medium properties -- such as self-interaction and coupling with standard model particles -- remains unknown, there are two main approaches to studying DM-admixed neutron stars. One approach involves considering non-gravitational interactions between ordinary matter (OM) and DM, such as the Higgs portal mechanism~\cite{Dutra:2022mxl, Lenzi:2022ypb, Hong:2024sey, Das:2018frc, Flores:2024hts, Das:2020vng}, and self-interacting DM models~\cite{Shirke:2023ktu, Thakur:2024btu}. In this scenario, the system can be effectively treated as a single fluid due to the non-gravitational interaction between ordinary matter and DM. The other approach involves ignoring the non-gravitational interaction between OM and DM, resulting in a two-fluid system where the different sectors interact solely through gravity. The two-fluid approach has been extensively studied (see e.g. Refs.~\cite{Collier:2022cpr,Shakeri:2022dwg, Miao:2022rqj, Emma:2022xjs, Hong:2024sey, Karkevandi:2021ygv, Ruter:2023uzc, Liu:2023ecz, Ivanytskyi:2019wxd, Buras-Stubbs:2024don, Rutherford:2022xeb,Thakur:2023aqm,Mahapatra:2024ywx,Thakur:2024mxs}).

The study of DM models in light of particle physics leads to stringent constraints on mass and DM couplings~\cite{Bauer:2017qwy}. In particular, the weakly interacting massive particle (WIMP) model stands out among other DM models because of the possibility of calculating the relic abundance directly from the weak interaction scale. This allows for the study of weak-scale DM particles in terrestrial laboratories~\cite{Kahlhoefer:2017dnp}. An indirect method gaining ground in the research community these days is the effect of DM on the observable properties of neutron stars~(NSs) such as mass-radius relation, tidal deformability, and gravitational wave signature. These effects have been extensively investigated using different DM and nuclear matter models~\cite{Leung:2022wcf, Kain:2021hpk, Leung:2011zz, das2022dark, Lenzi:2022ypb, Lourenco:2022fmf, Lourenco:2021dvh, xiang2014effects, Thakur:2023aqm, Thakur:2024scc}. For instance, self-annihilating DM inside NSs tends to heat it, thereby impacting its cooling properties during its evolution~\cite{Kouvaris:2007ay, Bertone:2007ae}. Non-self-annihilating DM, such as asymmetric DM~\cite{Kouvaris:2010jy} and mirror DM models~\cite{Ciarcelluti:2010ji, xiang2014effects}, accumulates inside NSs, influencing their macroscopic structure, tidal deformability, oscillatory properties, and gravitational wave signatures. These observable properties could provide an indirect method for detecting DM within compact stars. The constraints on different DM models are set by adjusting the model parameters against the observable properties of NSs. 

It has also been argued that DM can be captured within a star if it loses its kinetic energy through scattering with nuclear matter in the star. When this process takes place over some time it can lead to the accumulation of DM inside the star~\cite{Raj:2017wrv, PhysRevD.40.3221, gould1990neuton, kouvaris2008wimp, kouvaris2010can, deLavallaz:2010wp}. The quantity of DM that can be trapped inside the compact object will depend on the properties of the DM, the type of compact object (NS or white dwarf), and its evolutionary history. For example, WIMPs can accumulate in an NS due to elastic scattering with nucleons. Since the density of OM in the NSs dominates, the energy lost by the DM during collisions can be significant. This enhances the capture of DM inside the compact object~\cite{Raj:2017wrv, Guver:2012ba, PEREZGARCIA20126}.

The DM admixed NSs (DANSs) can be described as a two-fluid system where OM and DM are coupled through gravity only. In this setup, one fluid consists entirely of OM, described by its own equation of state (EoS), while the other fluid consists entirely of DM, described by a separate EoS. Therefore, the stellar properties such as the mass and radius are determined by solving the two-fluid Tolman-Oppenheiman-Volkoff (TOV) equations~\cite{sandin2009effects, 10.1143/PTP.47.444, Ciarcelluti:2010ji, de2010neutron}. This approach is not only theoretically suitable, but it is also straightforward to extend it to investigate the dynamical properties of the star in a self-consistent manner within the framework of general relativity. It also serves as the foundation for studying the stellar structure and stability of DANSs. Given that null results from DM direct detection experiments~\cite{LUX:2017ree, PandaX-II:2017hlx, XENON:2018voc} (see a general review in Refs.~\cite{Billard:2021uyg, Schumann:2019eaa}) have imposed strong constraints on the DM-nucleon coupling strengths. From the perspective of the TOV equations, the coupling strengths are effectively negligible~\cite{Gresham:2018rqo, Nelson:2018xtr}. Consequently, DANSs can be considered as two-fluid systems where inter-fluid interactions are purely gravitational.

This paper investigates the effect of DM on the evolution of supernova remnants from core birth as a neutrino-rich proto-neutron star (PNS) to maturity as neutrino-poor, cold-catalyzed NS~\cite{Raduta:2020fdn, Pons:1998mm, Janka:2012wk, Malfatti:2019tpg, Prakash:1996xs}. We use the quasi-static approximation, which allows us to assume a spherically symmetric PNS in hydrostatic equilibrium. The stellar evolution is then examined through the evolution of the intensive thermodynamic properties, such as entropy density and lepton fractions, over the Kelvin-Helmholtz timescale~\cite{1986ApJ...307..178B, Pons:1998mm, Roberts_2012}. The equations of state of DM and OM are calculated separately under the same thermodynamic conditions, assuming that DM and OM have the same entropy and lepton fraction. 

{However, to the best of our knowledge, at the time of writing this paper, there is no established reference on how DM thermalizes with OM in a way that would justify thermal equilibrium between them.  Nonetheless, since we are considering a mirrored fermionic DM model capable of forming isolated dark NSs, we adopt the same entropy per baryon as a reasonable approximation. Given that the interaction between the dark and visible sectors is purely gravitational, thermal equilibrium between the two components is unlikely. Nevertheless, by analogy with the visible sector, where NSs are expected to establish a well-defined entropy distribution, we extend this assumption to the dark sector as a first approximation (see Ref.~\cite{Bertone:2004pz} on thermalization and interaction of DM with OM using the standard cosmological model).} The effect of DM on the macroscopic and microscopic properties of the PNS is then investigated using the two-fluid formalism, in which DM and nuclear matter interact only gravitationally. This formalism allows us to introduce a fixed DM mass fraction into stellar matter and investigate its effect. 

The OM is composed of nucleons, hyperons, and leptons while the DM, on the other hand, comprises self-interacting dark fermions of the same kind (dark protons and dark neutrons). We use the relativistic mean-field approximation adjusted by the density-dependent meson-nucleon couplings (DDME2)~\cite{PhysRevC.71.024312} for OM and DM species at $\beta$-equilibrium. The coupling for the hyperons is taken from ~\cite{Lopes:2022vjx}, which extends the DDME2 to include hyperons and $\Delta$-resonances couplings based on symmetry group arguments. The main focus of this work is to examine how the presence of DM influences temperature distributions, particle distributions, and sound speed within the star throughout its evolution, assuming gravitational interactions. The structure of the resulting star is modeled, and its tidal deformability is calculated to compare the results with observable constraints.

The paper is organized as follows: Sec.~\ref{mp}, discusses the microphysics of the study which is divided into two parts. In Sec.~\ref{nm} we present the nuclear matter model and discuss its related parameterization. In Sec.~\ref{dm} we introduce the mirror DM model by focusing on the EoS at finite temperature. In Sec.~\ref{formalishm}, we present the two-fluid relativistic formalism that describes the transition from the microscopic to the macroscopic depiction of the DANSs. We present our findings and discuss them in Sec.~\ref{results}. The final remarks are presented in Sec.~\ref{fm}.

\section{Microphysics}\label{mp}

\subsection{Nuclear Matter Model}\label{nm}

The behavior of OM is governed by a relativistic mean-field model adjusted by density-dependent meson-nucleon couplings~\cite{Typel1999, PhysRevC.71.024312}, where the exchange of heavy mesons between baryons simulates the strong nuclear force acting between the constituent particles. The interaction is mediated by the non-linear mesons $\sigma$ (scalar), $\rho$ (isovector-vector), $\omega$ and $\phi$ (both vector-isoscalar, with $\phi$ carrying a hidden strangeness), described by the Lagrangian density
\begin{equation}
     \mathcal{L}_{\rm OM}= \mathcal{L}_{\rm H}+ \mathcal{L}_{\rm m}+ \mathcal{L}_{\rm L}.
               \end{equation}
Explicitly, the spin-1/2 baryon octet particles are represented by
\begin{align}\label{a}
 \mathcal{L}_{\rm H}= {}& \sum_{b}  \bar \psi_b \Big[  i \gamma^\mu\partial_\mu - \gamma^0  \big(g_{\omega b} \omega_0  +  g_{\phi b} \phi_0+ g_{\rho b} I_{3b} \rho_{03}  \big)\nonumber \\
 &- \Big( m_b- g_{\sigma b} \sigma_0 \Big)  \Big] \psi_b,
\end{align}
where $\psi_b$ represents the baryonic field of the baryon $b$ that can be either nucleons or hyperons, $I_{3b}$ is the isospin projection, and $m_b$ is the baryon mass. It is worth noting that the presence of hyperons is considered because neutrons, being fermions subject to the Pauli exclusion principle, make it energetically favorable to convert some of them into hyperons at high densities. The onset of the hyperons, particles with strangeness content, may lead to the softening of the EoS thereby reducing the NS maximum mass, particularly lower than the current observed mass of  $\sim 2M_\odot$. The challenge of reconciling the presence of hyperons in the NS interior with the current observational mass is an open problem under active research in nuclear astrophysics. This problem is known in the literature as the `hyperon puzzle' \cite{Bombaci:2016xzl}. The mesonic part is given by
\begin{align}\label{a11}
 \mathcal{L}_{\rm m}= - \frac{1}{2} m_\sigma^2 \sigma_0^2  +\frac{1}{2} m_\omega^2 \omega_0^2  +\frac{1}{2} m_\phi^2 \phi_0^2 +\frac{1}{2} m_\rho^2 \rho_{03}^2,
\end{align}
where $m_i$ ($i=\{\sigma,\omega,\rho,\phi\}$) is the meson mass. The free leptons that are introduced in the stellar matter to ensure charge neutrality in the stellar matter are described by the Dirac-like Lagrangian,
\begin{equation}\label{l1}
    \mathcal{L}_{\rm L} = \sum_L\Bar{\psi}_L\left(i\gamma^\mu\partial_\mu-m_L\right)\psi_L,
\end{equation}
where $\psi_L$ is the lepton fields and the subscript $L$ accounts for all the leptons present in the stellar matter.
The subscript `$0$' represents the mean-field approximation of the fields.

The leptons and baryons are spin-1/2 particles, each with a degeneracy of two. However, when the star traps neutrinos at finite temperatures, we consider electron neutrinos ($\nu_e$) with a degeneracy of one. At this stage, the muons are not considered since their presence becomes relevant when the star is neutrino-free per supernova physics~\cite{Malfatti:2019tpg, Prakash:1996xs}. For a complete analysis of stellar evolution, we consider $e$ and  $\nu_e$ along with baryons in the early stages of the star's life. In the later stages, after neutrino diffusion, we consider $e$ and $\mu$, while tau leptons ($\tau$) are assumed to be too heavy to be present~\cite{Pons:1998mm, Raduta:2020fdn}.

We employed the DDME2 parameterization~\cite{Typel1999,PhysRevC.71.024312} with density-dependent meson-nucleon coupling defined as,
\begin{equation}\label{cp}
    g_{i b} (n_B) = g_{ib} (n_0)a_i  \frac{1+b_i (\eta + d_i)^2}{1 +c_i (\eta + d_i)^2},
\end{equation}
where $n_B$ is the total baryon density, for $i=\sigma, \omega, \phi$, also, 
\begin{equation}
    g_{\rho b} (n_B) = g_{\rho b} (n_0) \exp\left[ - a_\rho \big( \eta -1 \big) \right],
\end{equation}
where $\eta=n_B/n_0$, with $n_0=0.152\,\rm fm^{-3}$ being the nuclear saturation density. From this particular parametrization, the following nuclear empirical parameters at $n_B=n_0$ are obtained: $E_B = -16.14$~MeV~(binding energy), $K_0 = 251.9$~MeV~(incompressibility), $J = 32.3$~MeV~(symmetry energy), and $L_0 =51.3$~MeV~(symmetry energy slope). These results are in good agreement with recent constraints on the properties of symmetric nuclear matter~\cite{Reed:2021nqk, Lattimer:2023rpe, Dutra:2014qga}. The value of the constants $a_i,\;b_i, \; c_i,\;d_i$ and the corresponding meson masses are displayed in Tab.~\ref{tab:param}.

An extension to include hyperons in the model is done relative to the meson-nucleon couplings ($\chi_{bi}=g_{ib}/g_{iN}$). There are different forms of obtaining these couplings in the literature~\cite{Glendenning:1991es, Lopes:2020rqn, Pais:1966eox, Weissenborn:2011kb}. Here we use the results reported in~\cite{Lopes:2022vjx}, where the authors determined the meson-baryon coupling covering hyperons and $\Delta$-resonances 
using SU(3) and SU(6) symmetry arguments, displayed in Tab.~\ref{tab:hyp}. The advantage of this coupling is that it helps go around the `hyperon puzzle' by obtaining NSs with maximum masses within the $2M_\odot$ threshold.

\begin{table}[!t]
\centering
\caption{DDME2 parameter set.}
\begin{tabular}{ |c| c| c| c| c| c| c| }
\hline
 meson($i$) & $m_i(\text{MeV})$ & $a_i$ & $b_i$ & $c_i$ & $d_i$ & $g_{i N} (n_0)$\\
 \hline
 $\sigma$ & 550.1238 & 1.3881 & 1.0943 & 1.7057 & 0.4421 & 10.5396 \\  
 $\omega$ & 783 & 1.3892 & 0.9240 & 1.4620 & 0.4775 & 13.0189  \\
 $\rho$ & 763 & 0.5647 & --- & --- & --- & 7.3672 \\
 \hline
\end{tabular}
\label{tab:param}

\tabcolsep=0.46cm
\centering
\caption{Ratio of the baryon coupling to the corresponding nucleon coupling for hyperons.}
\begin{tabular}{ |c | c| c| c| c| } 
\hline
b & $\chi_{\omega b}$ & $\chi_{\sigma b}$ & $\chi_{\rho b}$ & $\chi_{\phi b}$  \\
\hline
$\Lambda$ & 0.714 & 0.650 & 0 & -0.808  \\  
$\Sigma^0$ & 1 & 0.735 & 0 & -0.404  \\
 $\Sigma^{-}$, $\Sigma^{+}$ & 1 & 0.735 & 0.5 & -0.404  \\
$\Xi^-$, $\Xi^0$  & 0.571 & 0.476 & 0 & -0.606 \\
\hline
\end{tabular}
\label{tab:hyp}
\end{table}

From the matter EoS, i.e. the $P_{OM}$ vs. $\varepsilon_{OM}$ relation, one can obtain the free energy relation, $\mathcal{F}_B = \varepsilon_{OM} - Ts$, which allows us to obtain the entropy density $s$ (where $s=s_Bn_B$, with $s_B$ the entropy per baryon and $n_B$ the total baryon density). The quantities \( P_{OM} \) and \( \varepsilon_{OM} \) are the respective total pressure and total energy density of the ordinary matter, encompassing all the baryons and leptons present in the system. For stellar matter, it reads
\begin{equation}
sT =  \varepsilon_{OM} + P_{OM} - 
\sum_b \mu_b n_b - \sum_L \mu_L n_L
,
\end{equation}
with the sum running over all considered particle species. $n_b$ is the baryon density, $\mu_b$ is the baryon chemical potential, $n_L$ is the lepton number density, $\mu_L$ is the lepton chemical potential and $T$ is temperature. In the neutrino-transparent stellar matter,
\begin{equation}\label{s1}
     sT=P_{OM}+\varepsilon_{OM} -n_B\mu_B,
 \end{equation}
 where $\mu_B$ is the total baryon chemical potential.
  In the neutrino-trapped stellar matter, this expression is modified by the neutrino number density $n_{\nu_e}$, electron number density $n_e$, and the neutrino chemical potential $\mu_{\nu_e}$, yielding
 \begin{equation}\label{s2}
    sT=P_{OM}+\varepsilon_{OM} -n_B\mu_B-\mu_{\nu_e}(n_{\nu_e} +n_e).
 \end{equation}
Equations~(\ref{s1}) and~(\ref{s2}) are the simplified versions of the $s_B$ expression using $\beta$-equilibrium and charge neutrality conditions. We calculate the temperature distributions along with the EoS by fixing $s_B$ together with the lepton fraction ($Y_{L,e}$) to control the neutrino concentration in the neutrino-trapped stage and $Y_{\nu_e} =0$ in the neutrino-free stage.
A detailed calculation of the EoS can be found in~\cite{Issifu:2023qyi} where we adopted the EoS for this work. In addition to that, density-dependent meson-nucleon coupling has been used to study PNSs with exotic baryons in~\cite{Sedrakian:2022kgj, Raduta:2020fdn, Malfatti:2019tpg, Issifu:2024fuw} where EoSs have been explicitly derived. 

PNSs are typically investigated at a fixed $s_B$ instead of $T$ because entropy provides a natural framework for describing the thermodynamic state of matter during evolution and ensures computational consistency in studying these objects \cite{Pons:1998mm}. During core collapse and bounce, the process is approximately adiabatic outside the shock, leading to a more uniform entropy distribution in different mass shells, which helps characterize the star’s thermodynamic state. Additionally, neutrino emission, which dominates the cooling processes of PNSs, depends strongly on $s_B$ and lepton fraction rather than temperature \cite{Camelio:2017nka, Janka:2016fox}. In contrast, temperature varies significantly with density in newborn NSs, with estimates suggesting it can reach 30–50 MeV in the core while the outer layers remain much cooler. Fixing entropy also facilitates comparisons with supernova simulations, which primarily track entropy evolution rather than temperature \cite{10.1093/mnras/sty2585}.

\subsection{Dark Matter Model}\label{dm}

 We adopt a mirror DM model in which the DM contains a mirrored sector that replicates the interactions observed in the visible sector, specifically mimicking the interactions of nuclear matter. 
We consider self-interacting fermionic DM particles ($\psi_D$) that interact with the OM sector through gravity. In the Lagrangian density that incorporates the dynamics of the fermionic DM, we assume a dark scalar meson $\Tilde{\sigma}$ that couples with the DM candidate through $g_{\tilde{\sigma}}{\overline\psi_D}\psi_{\rm D}\tilde{\sigma}$ and a dark vector meson ($V^\mu$) that couples to the conserved DM current by $g_v{\overline\psi_D}\gamma_{\mu}\psi_{\rm D}V^{\rm \mu}$~\cite{xiang2014effects}. A similar model was used in~\cite{Thakur:2023aqm} where the authors considered only vector interactions. The Lagrangian density is given by
\begin{align}\label{L1}
    {\cal L}_{\rm DM}={}&{\overline\psi_D}[ (i\gamma_\mu\partial^\mu- \gamma^0g_{v} V_0)-(m_D -g_{\tilde{\sigma}} \tilde{\sigma}_0)]\psi_D \nonumber\\
    -&\frac{1}{2}m_{\rm \tilde{\sigma}}^{2}\tilde{\sigma}_0^{2}
    +\frac{1}{2}m_{v}^{2} V_0^2,
\end{align}
where $m_D$ is the mass of the dark fermion, $\tilde{\sigma}_0$ and $V_0$ are the RMF representations of $\tilde{\sigma}$ and $V^\mu$ respectively.

This Lagrangian density is similar to the Walecka model for nucleon (see Ref.~\cite{Menezes:2021jmw} and references therein), where we have omitted the isospin projection that differentiates between proton and neutron, thus, referring to {dark protons and dark neutrons} as identical dark fermions. To calculate the EoS for the model, we use the mean-field approximation with density-dependent coupling similar to the approach presented in Sec.~\ref{nm} for the OM. Therefore, we replaced $g_{\tilde{\sigma}}$ and $g_v$ with $g_{\sigma N}$ and $g_{\omega N}$ and the meson masses $m_{\rm \tilde{\sigma}}$ and $m_{v}$ with $m_\sigma$ and $m_\omega$ respectively, following Tab.~\ref{tab:param}, and $m_D = 939$~MeV. This formalism mirrors the visible sector by adopting the same saturation density, coupling constants, and meson masses as nuclear matter. This assumption reduces the arbitrariness in fixing the couplings. The coupling is defined in the same form as Eq.~(\ref{cp}), with $\eta \rightarrow n_{B'}/n_0$ ($n_{B'}$ is the total dark baryon density), using the fit parameters on Tab.~\ref{tab:param}.  The detailed calculation of the EoS of the above Lagrangian density at $T=0$ can be found in~\cite{xiang2014effects}.

{At finite temperature, the respective dark energy density $\varepsilon_D$ becomes 
\begin{align}
     \varepsilon_D={}& \sum_{D}\left(\gamma_D \int \frac{d^3 k_D}{( 2\pi)^3} E_D \left [f_{D+} +f_{D-} \right]+\frac{g_{\rm \omega N}^2}{2 m_{\rm \omega}^2}n_D^2 \right)
\nonumber\\
&+\frac{m_{\rm \sigma}^2}{2 g_{\rm \sigma N}^2} (m_D-m_D ^*)^2 \nonumber\\&+ \sum_{D_L}\gamma_{D_L} \int \frac{d^3 k_{D_L}}{( 2\pi)^3} E_{D_L} \left [f_{{D_L}+} +f_{{D_L}-} \right],
\end{align}
where $k_D$ is the momentum, and $m_D ^*$ is the effective mass, which is given by 
\begin{equation}
m_D^* = m_D - g_{\rm \sigma N}\tilde{\sigma}_0.
\end{equation}
Subscript $D_L$ represents dark leptons with $\sum_{D_L}$ representing the summation over all the dark leptons in the matter. $\gamma_{D_L}$ is also the degeneracy of the dark lepton, $\gamma_{D_L}=2$ for dark electrons and dark muons and $\gamma_{D_L}=1$ for dark neutrinos similar to the visible sector. The chemical potential and the mass of the dark leptons are constants. The properties of the dark leptons are mirrored from the visible sector using the noninteracting Dirac-like Lagrangian as presented in Eq.~\eqref{l1}.  The single particle energy is given by $E_D = \sqrt{{k_D}^2+(m_D^*)^2}$ and $E_{D_L} = \sqrt{{k_{D_L}}^2+(m_{D_L})^2}$ for the DM particles and the dark leptons respectively, with $m_{D_L}$ the mass of the dark lepton. The pressure, on the other hand, is given by
\begin{equation}
    P_D = n_D^2\dfrac{\partial}{\partial n_D}\left(\dfrac{\varepsilon_D}{n_D}\right),
\end{equation}
which yields a correction term, 
\begin{equation}
    P_r^D = n_D\Sigma^r_D,
\end{equation}
to ensure the thermodynamic consistency of the model. The rearrangement term $\Sigma_D^r$, is given explicitly as
\begin{equation}
    \Sigma^r_D =  \frac{\partial g_{\omega N}}{\partial n_D} V_0 n_D  - \frac{\partial g_{\sigma N}}{\partial n_D} \tilde{\sigma}_0 n_D^s,
\end{equation}
where $n_D^s$ is the scalar density and $n_D$ is the DM density. The respective $n_D$ and $n_D^s$ are given by
\begin{equation}
n_D = \sum_{D}\gamma_D \int \frac{d^3 k_D}{(2\pi)^3}  \left[f_{D\,+} - f_{D\,-}  \right],
\end{equation}
and 
\begin{equation}
    n_{D}^s =\sum_{D}\gamma_D \int \frac{d^3 k_D}{(2\pi)^3} \frac{m^\ast_D}{E_D} \left[f_{D\,+} + f_{D\,-}  \right].
\end{equation}
where 
\begin{equation}\label{fd}
    f_{D \pm}(k_D) = \frac{1}{1+\exp[(E_D \mp \mu_{D}^\ast)/T_D]}, \nonumber
\end{equation}
is the Fermi distribution function with the effective chemical potential 
\begin{equation}
    \mu_{p', n'}^\ast = \mu_{p', n'}- g_{\omega N} V_0  - \Sigma^r_D,
\end{equation}
where $\mu_{p', n'}$ is the real chemical potential of the dark protons ($p'$) and dark neutrons ($n'$) and $T_D$ is temperature. We calculate the EoS at a fixed dark entropy $\tilde{s}$ ($\tilde{s}= s_Dn_D$, where $\tilde{s}$ is DM entropy density, $s_D$ is the total entropy per DM particle) using the free energy expression $\mathcal{F}_D = \varepsilon_D - T_D\tilde{s}$ for dark particles.  Similarly to the visible sector, the dark sector is treated as a single fermionic fluid governed by the Pauli exclusion principle. Therefore, a balance between dark fermions and dark leptons must be maintained through a dark $\beta$-equilibrium reaction to accurately characterize the physical properties of the dark stellar system. Since we have identical dark fermions, in the neutrino-trapped matter 
{\begin{equation}
    \mu_{B'} = \mu_{p'} + \mu_{e'}-\mu_{\nu_e'}, \quad{\rm with}\quad \mu_{n'} = \mu_{B'}
\end{equation}
where $\mu_{e'}$ is the chemical potential of the dark electrons and $\mu_{\nu_e'}$ is the chemical potential of the dark electron neutrinos, and in the neutrino transparent matter 
\begin{equation}
    \mu_{B'} = \mu_{p'}+\mu_{e'},
\end{equation}
with $\mu_{B'}$ the total baryon density of the dark sector. The dark muons become relevant when the matter is neutrino-transparent as mentioned above, however, their chemical potential is the same as that of electrons ($\mu_{e'}=\mu_{\mu'}$). In the mirrored model scenario, DM in isolation can form compact objects (a dark NS, perhaps), with a simpler composition, assuming the presence of only dark nucleons. Since the dark NSs are physically observable objects in the dark world, we impose dark charge neutrality like in the visible sector:
\begin{equation}
    n_{p'} = n_{e'},
\end{equation}
for neutrino-trapped matter, and 
\begin{equation}
    n_{p'} = n_{e'}+n_{\mu'},
\end{equation}
for neutrino-transparent matter, where $n_{e'}$ and $n_{\mu'}$ are respectively dark electron and dark muon number densities. Using the dark $\beta$-equilibruim together with dark charge neutrality relations, the $\mathcal{F}_D$ is modified through the expressions
\begin{equation}
     \tilde{s}T_D=P_D+\varepsilon_D - n_{B'}\mu_{B'},
\end{equation}
in neutrino-transparent matter and 
\begin{equation}
    \tilde{s}T_D=P_D+\varepsilon_D -n_{B'}\mu_{B'}-\mu_{\nu_e'}(n_{\nu_e'} +n_{e'}),
\end{equation}
in neutrino-trapped matter.} Studies related to mirror dark matter models and their applications to compact astrophysical objects can be found in~\cite{Sandin:2008db, Foot:2004pa, Berezhiani:2005ek, Hippert:2021fch}.

\section{Two Fluid Formalism \label{formalishm}}
We have employed a two-fluid TOV formalism to analyze the structure of NSs with a mixture of DM, referred to as 
DANSs~\cite{Das:2020ecp}. DM and OM are treated separately within this framework and interact solely through gravitational interaction. Consequently, each fluid follows its conservation of energy-momentum tensor.
The TOV equations governing the behavior of this two-fluid system are given by~\cite{Das:2020ecp, Rutherford:2022xeb}
\begin{align}
\frac{dP_{OM}}{dr} ={}& -(P_{OM} + \varepsilon_{OM})\frac{4\pi r^3(P_{OM}+P_{D})+M(r)}{r(r-2M(r))} \label{tov},\\
\frac{dP_{D}}{dr} ={}& -(P_{D} + \varepsilon_{D})\frac{4\pi r^3(P_{OM}+P_{D})+M(r)}{r(r-2M(r))}\label{tov1},
\end{align}
and
\begin{equation}\label{tov2}
\frac{dM(r)}{dr} = 4\pi (\varepsilon_{OM} + \varepsilon_D) r^2.
\end{equation}
When investigating the influence of DM on NSs, it is useful to define a DM mass fraction $F_{D}$~\cite{Rutherford:2022xeb}, defined as
\begin{equation}
F_{D} = \frac{M_{D}(R_D)}{M{(R)}}.
\end{equation}
Here, $M_{D}(R_{D})=4\pi\int_0^{R_D} r^2\varepsilon_{D} (r) dr$ represents the total accumulated DM gravitational mass within $R_{D}$, where the DM pressure reaches zero. The gravitational mass of a star composed solely of ordinary matter follows a similar procedure. $M(R)$ represents the total mass (a combined mass of the DM and the OM) enclosed within a radius $R$. Here, $R$ represents the outer radius of the star, which, in this case, corresponds to the radius of the OM, as the current study focuses on the presence of DM in the core of the NS. Based on the DM mass fraction, it is possible to determine how much the gravitational mass of the DANS contributes to the star's total mass.

Besides mass and radius, NSs' tidal deformability is crucial in their structural characteristics. The tidal gravitational field generated by their companion causes the two NSs in a binary NS system to undergo quadrupole deformations during the final inspiral stages. 
As a result of the tidal forces exerted by the partner star of an NS, the magnitude of the deformation that occurs is described as tidal deformability, which quantifies the extent to which it distorts under those forces.

The dimensionless  tidal deformability is defined as 
\begin{equation}\label{td}
    \Lambda = 2/3 ~ k_2 ~ C^{-5},
\end{equation}
where $C= M/R$ and $k_2$ are known as the compactness and Love number of the deformed star respectively. The value of $k_2$ for the two-fluid system can be obtained by solving the differential equation for radial perturbation, 
\begin{equation}
 r \frac{d y(r)}{dr} + {y(r)}^2 + y(r) F(r) + r^2 Q(r) = 0,
  \end{equation}
where,
\begin{equation}
    F(r) = \frac{r-4 \pi r^3 \left\{ [\varepsilon_{OM}(r)+\varepsilon_D(r)] - [P_{OM}(r)+P_D(r)]\right\} }{r-2
M(r)},
\end{equation}
and
\begin{widetext}
\vspace{-4mm}
\begin{align}
Q(r) ={}& \frac{4 \pi r \left\{5 [\varepsilon_{OM}(r)+\varepsilon_D(r)] +9 [P_{OM}(r)+P_D(r)] +
\frac{\varepsilon_{OM}(r) + P_{OM}(r)}{\partial P_{OM}(r)/\partial
\varepsilon_{OM}(r)}+\frac{\varepsilon_D(r) + P_D(r)}{\partial P_D(r)/\partial
\varepsilon_D(r)} - \frac{6}{4 \pi r^2}\right\}}{r-2M(r)} \nonumber  \\
&- 4\left\{\frac{M(r) + 4 \pi r^3
[P_{OM}(r)+P_D(r)]}{r^2\left(1-2M(r)/r\right)}\right\}^2, 
\end{align}  
\end{widetext}
together with the two-fluid TOV equations with the required boundary conditions~\cite{Das:2020ecp, Karkevandi:2021ygv}. 

\section{Results} 
\label{results}

\begin{table*}[!t]
\small
\centering
\caption{Maximum stellar mass ($M_{\rm max}$), its radius ($R_{\rm max}$), the radii of the canonical and $\rm M=2.1 M_\odot$ stars ($\rm R_{1.4}$ and $\rm R_{2.1}$), the canonical star tidal deformability ($\Lambda_{1.4}$), and baryonic mass ($M_{\rm B}$) for different configurations, OM compositions, and proportions of DM mass fraction in an NS.}
\tabcolsep=0.3cm
\begin{tabular}{|c|cccccccc|}
\hline
Configuration                             & Composition         & DM {[}\%{]} & $M_{\rm max}$ {[}$\rm M_\odot${]} & $R_{\rm max}$ {[}km{]} & $R_{1.4}$ {[}km{]} & $R_{2.1}$ {[}km{]} & $\Lambda_{1.4}$ & $M_{\rm B_{2.10}}$ {[}$\rm M_\odot${]} \\ \hline
\multirow{6}{*}{$s_B=1$; $Y_{L,e}=0.4$}   & \multirow{3}{*}{N}  & 0           & 2.43                         & 12.49                  & 15.41 & 14.15             & 865             & 2.33        \\
                                          &                     & 1           & 2.40                         & 12.40                  & 15.37     & 14.03         & 704             & 2.28            \\
                                          &                     & 5           & 2.30                         & 12.12                  & 14.85     & 13.47         & 480             & 2.23           \\ \cline{2-9} 
                                          & \multirow{3}{*}{NH} & 0           & 2.32                         & 12.53                  & 15.41      & 14.04        & 863             & 2.33            \\
                                          &                     & 1           & 2.29                         & 12.46                  & 15.37       & 13.15       & 708             & 2.28           \\
                                          &                     & 5           & 2.19                         & 12.12                  & 14.86    & 13.88          & 483             & 2.23            \\ \hline
\multirow{6}{*}{$s_B=2$; $Y_{L,e}=0.2$}   & \multirow{3}{*}{N}  & 0           & 2.48                         & 13.01                  & 16.62      & 15.03        & 1281            & 2.34           \\
                                          &                     & 1           & 2.45                         & 12.93                  & 16.57     & 14.89         & 903             & 2.29           \\
                                          &                     & 5           & 2.35                         & 12.64                  & 15.96      & 14.50        & 596             & 2.25            \\ \cline{2-9} 
                                          & \multirow{3}{*}{NH} & 0           & 2.30                         & 12.72                  & 16.43   &    14.42       & 863             & 2.35            \\
                                          &                     & 1           & 2.28                         & 12.65                  & 16.36    & 14.24          & 708             & 2.29          \\
                                          &                     & 5           & 2.18                         & 12.31                  & 15.74     & 13.08         & 483             & 2.25           \\ \hline
\multirow{6}{*}{$s_B=2$; $Y_{\nu_e}=0.0$} & \multirow{3}{*}{N}  & 0           & 2.49                         & 13.07                  & 16.97         & 15.18     & 1299            & 2.35          \\
                                          &                     & 1           & 2.46                         & 13.02                  & 16.95      & 15.05        & 1036            & 2.27           \\
                                          &                     & 5           & 2.34                         & 12.71                  & 16.31      & 13.82        & 753             & 2.25           \\ \cline{2-9} 
                                          & \multirow{3}{*}{NH} & 0           & 2.28                         & 12.70                  & 16.67      & 14.36        & 1152            & 2.37            \\
                                          &                     & 1           & 2.25                         & 12.65                  & 16.63    & 14.15          & 988             & 2.29            \\
                                          &                     & 5           & 2.14                         & 12.33                  & 15.96    & 13.13          & 737             & 2.27           \\ \hline
\multirow{6}{*}{$T=0$}                    & \multirow{3}{*}{N}  & 0           & 2.48                         & 12.06                  & 13.19      & 13.20        & 704             & 2.44            \\
                                          &                     & 1           & 2.45                         & 11.94                  & 13.11     &13.08         & 500             & 2.40            \\
                                          &                     & 5           & 2.33                         & 11.66                  & 12.76    &   12.61     & 356             & 2.36           \\ \cline{2-9} 
                                          & \multirow{3}{*}{NH} & 0           & 2.26                         & 11.94                  & 13.19      & 12.98        & 704             & 2.35            \\
                                          &                     & 1           & 2.22                         & 11.82                  & 13.12     & 12.80         & 490             & 2.27            \\
                                          &                     & 5           & 2.11                         & 11.48                  & 12.75      & 11.83        & 348             & 2.24            \\ \hline
\end{tabular}
\label{SP}
\end{table*}

\begin{figure}[!t]
\centering
\includegraphics[width=\linewidth]{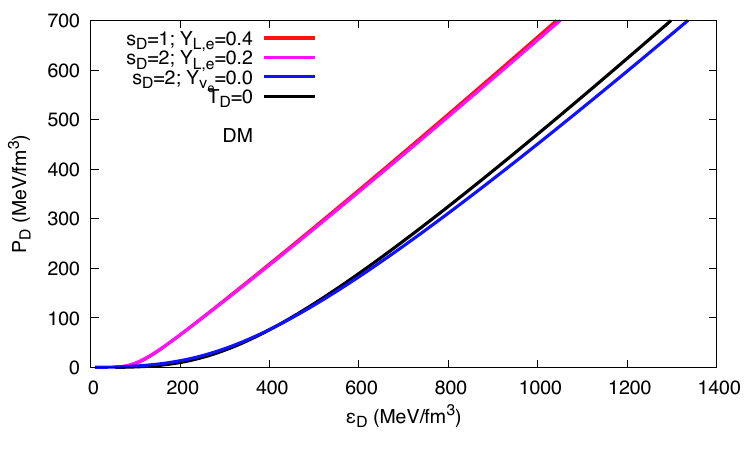}\\
\includegraphics[width=\linewidth]{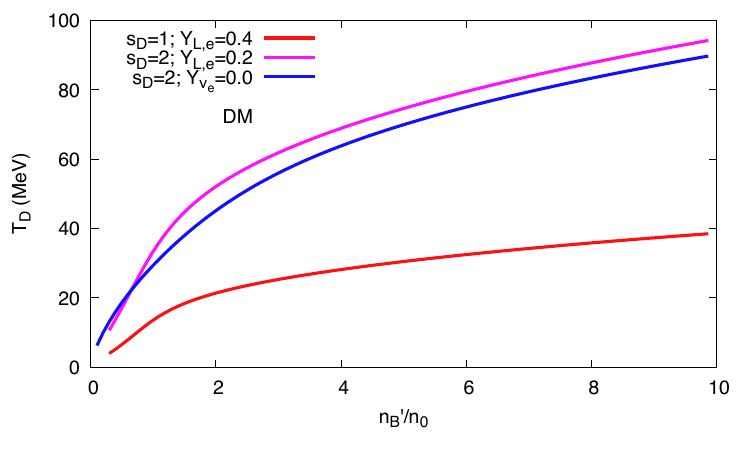}
\caption{EoS for dark matter configurations with and without trapped neutrinos (top), and the corresponding temperature distributions (bottom).
}
\label{fig:pxe}
\end{figure}

The OM EoS adopted in this work along with its temperature profiles can be found in~\cite{Issifu:2023qyi}. Additionally, this EoS was also applied in~\cite{Issifu:2024fuw} to study PNSs with quark cores. In Fig.~\ref{fig:pxe}, we present the EoS of the DM and its corresponding temperature profiles against DM density. 
The neutrino-rich dark PNSs at birth, when they are trapping neutrinos, correspond to stiffer EoS and lower temperature profiles. As the dark neutrinos diffuse out of the star, the star heats up, and its EoS softens. This process continues until the epoch of maximum heating, at which point all the neutrinos have escaped ($s_D=2,\; Y_{\nu_e}=0$), and the star becomes neutrino-transparent. At this stage, the star continues to cool through thermal radiation until it reaches the state of a cold-catalyzed compact dark star at $T=0$. The observed behavior of the dark PNSs is qualitatively similar to what is observed in the visible sector~\cite{Pons:1998mm, Raduta:2020fdn, Malfatti:2019tpg, Prakash:1996xs}. 

\begin{figure}[!t]
\centering
\includegraphics[width=\linewidth]{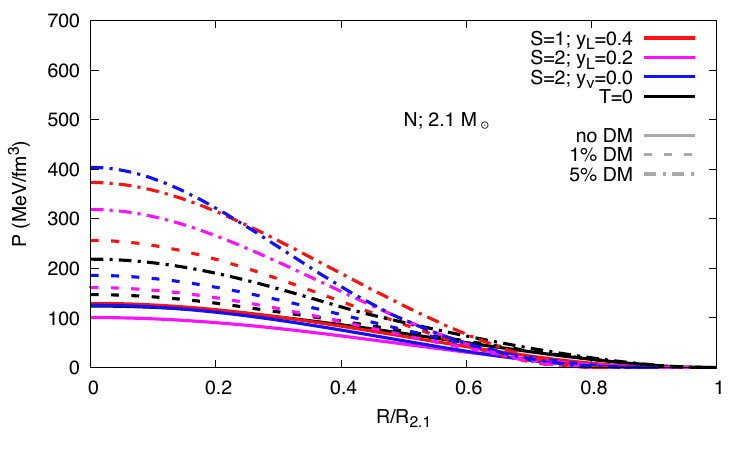}\\
\includegraphics[width=\linewidth]{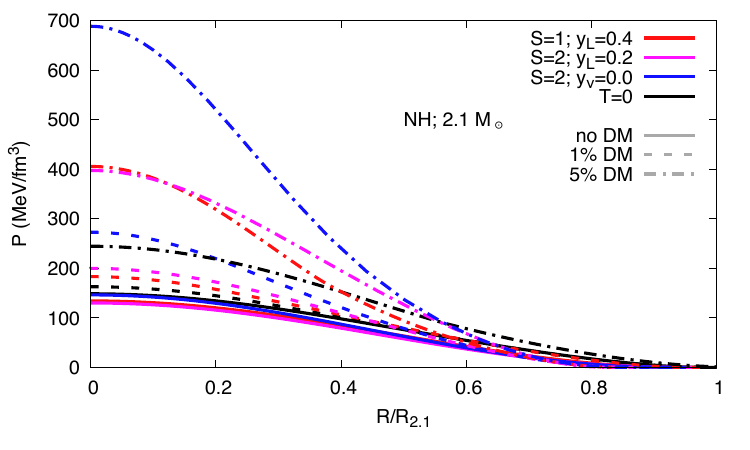}
\caption{The pressure profile of $ 2.1M_\odot$ stellar configuration, considering nucleonic (top) and hyperonic (bottom) admix DM.}
\label{fig:rxp}
\end{figure}

{Determining the microscopic properties of DANSs, such as particle fractions and temperature distributions, as well as macroscopic thermodynamic quantities like the speed of sound, is inherently complex. This complexity arises from the need to account for gravitational effects, which can only be properly incorporated through the two-fluid TOV equations. In this work, we employ numerical techniques that help us to estimate these quantities without neglecting the gravitational effect. Considering that the stellar configurations are in hydrostatic equilibrium, satisfying the two-fluid TOV equations, the pressure gradient balances the gravitational force at each point inside the star. Thus, the pressure at each point is the sum of the pressures from the two fluids: $P=P_{OM}+P_D$. The total mass of the star on the other hand, is determined by integrating over the total energy density, which is the sum of the energy densities of the two fluids: $\varepsilon = \varepsilon_{OM}+\varepsilon_D$, as represented in the TOV equations (\ref{tov1}) to (\ref{tov2}). This approach is particularly useful because the total pressure profile $P(r)$ is derived from the two-fluid TOV equations, which naturally incorporate the effects of gravity on the net pressure and central energy density distribution. The relationship between $P$, $P_{OM}$, and normalized $R$ is shown in Fig.~\ref{fig:rxp}. The plot shows that the emergence of new degrees of freedom and increased DM mass fraction increase the pressure in the stellar core.

To estimate the quantities, we calculate the pressure profile of a star with a specific gravitational mass, in this case $2.10 M_\odot$. The pressure profile is determined by numerically integrating the two-fluid TOV equations from the star's center ($r=0$) to its surface ($r=R$). Then we extract the pressure, central energy density, and radius corresponding to this star. The integration is performed using the boundary conditions: $P(r=0)=P_{\rm center}$, $M(r=0)=0$, $P(r=R)=0$ and $M(R)= 2.10 M_\odot$ within the two-fluid TOV framework. For each particle fraction or temperature distribution corresponding to $P_{OM}$, we determine its equivalent in $P$ using the cubic spline interpolation method. This serves as a reasonable approximation in the absence of a rigorous first-principles approach that fully accounts for gravity.}

\begin{figure*}[!t]
\centering
\includegraphics[width=0.45\linewidth]{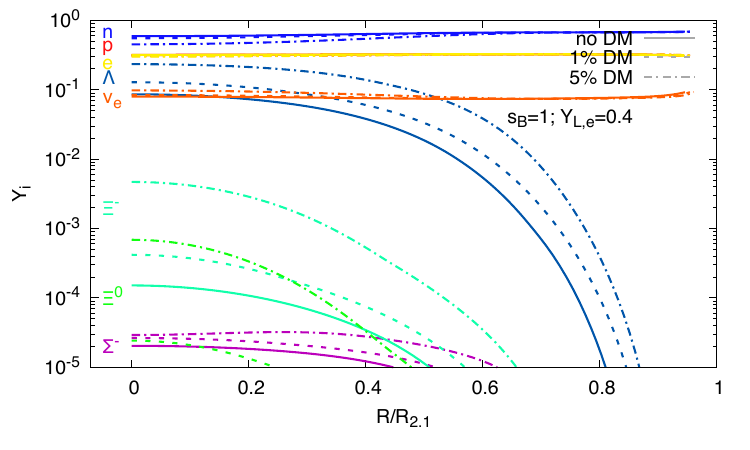}
\includegraphics[width=0.45\linewidth]{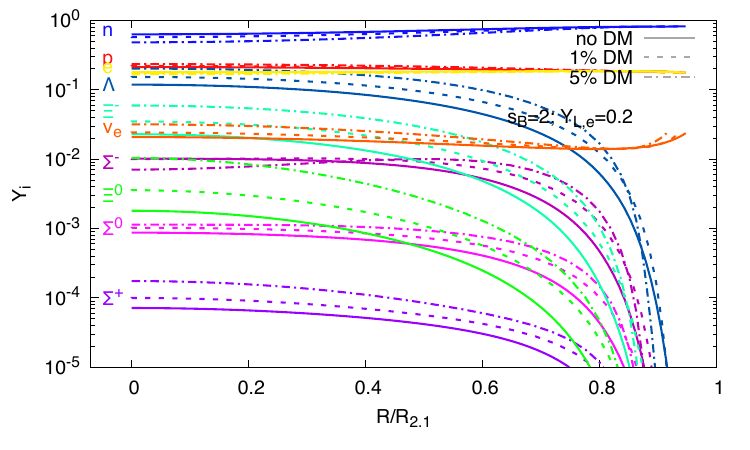}
\includegraphics[width=0.45\linewidth]{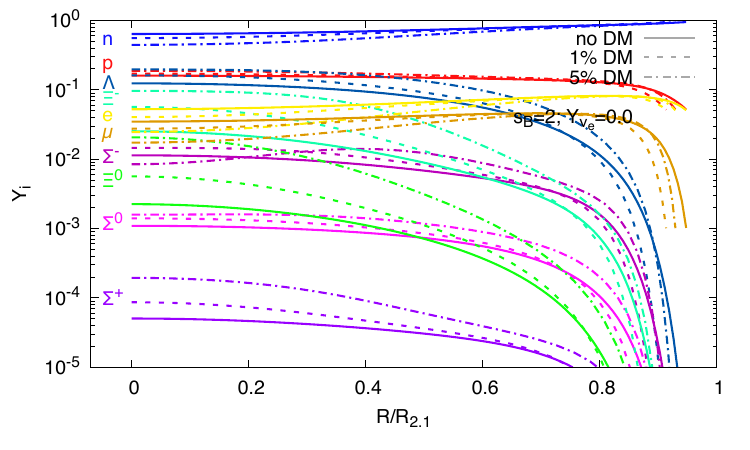} 
\includegraphics[width=0.45\linewidth]{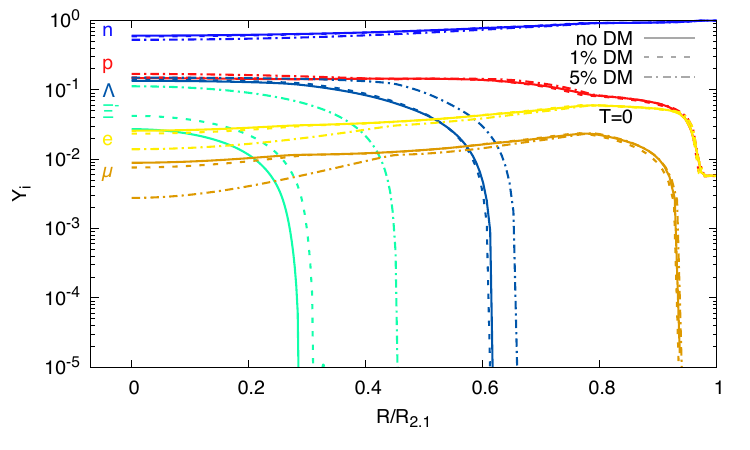}
\caption{Particle population profile of a $2.1M_\odot$ nucleon plus hyperon stars for different content of DM and lepton fractions. The curves are represented as follows: solid lines correspond to no DM mass fraction, dashed lines indicate 1\% DM mass fraction and dash-dot lines represent 5\% DM mass fraction  (indicated on the top- 
 right corner).  Top-left: $s_B=1$ and top-right: $s_B=2$ and $Y_{L,e}=0.2$;  bottom-left: $s_B=2$ and $Y_{L,e}=0$; bottom-right: $T=0$ and $Y_{L,e}=0$. 
}
\label{fig:yxr}
\end{figure*}
The particle distribution ($Y_i$) is calculated from
\begin{equation}
Y_i= \dfrac{n_i}{n_B},
\end{equation}
$i$ represents the different particles in the stellar matter and $n$ is their number densities. Fig.~\ref{fig:yxr} shows the particle distributions within a star with mass $2.1M_\odot$ at different stages of the stellar evolution. Along the panels, we present snapshots of the stellar evolution from core birth to the stages where the star heats up by neutrino diffusion and then to `maturity' as cold-catalyzed NS several years after the supernova explosion. The first panel (top-left) represents the ambient conditions of core birth when the star is neutrino-rich. The second panel (top-right) represents the period of deleptonization after the supernova explosion, at this stage the star is about 0.5 to 1 second old. Here, the stellar matter heats up due to neutrino diffusion, so its temperature is higher than in the first stage. The third stage (bottom-left) occurs when neutrinos fully escape from the stellar core. At this stage, the star reaches its maximum temperature and then begins to cool, roughly 10 to 15 seconds after the core birth. The last panel (bottom-right) is when the star has cooled down through neutrino diffusion and thermal radiation to form a cold-catalyzed NS, over several years. 
It is important to mention that two fates await the PNS after the supernova explosion: If the pressure generated by the supernova explosion is insufficient to eject the outer mantle of the progenitor massive star completely, the PNS may continue to accrete matter, potentially leading to the formation of a black hole. In this case, it is believed that the neutrino emission can abruptly stop~\cite{walk2020neutrino} and the evolution of the PNS terminates without moving to the second stage. On the other hand, if the pressure generated by the explosion is sufficient to lift the surrounding stellar envelope and induce deleptonization, the outer mantle collapses, allowing the PNS to continue its evolution. During deleptonization, transitioning from the second to the third panel, the likelihood of strangeness-rich particles emerging in the stellar matter increases. This can abruptly soften the EoS, potentially forming a black hole if the gravitational pressure is sufficiently strong~\cite{Prakash:1996xs}. Therefore, Fig.~\ref{fig:yxr} is constructed assuming that the PNS continues to evolve without collapsing into a black hole.

From the top to the bottom panels, we observe that the presence of DM influences the particle fractions depending on the effect of pressure on their production. Specifically, DM enhances the fractions of particles whose production is favored by increased pressure, as a consequence the presence of DM favors the increase in the quantity of all species of hyperons ($\Lambda,\; \Xi,\; \Sigma$), in particular, the $\Xi$ are the ones that show a higher increase due to the presence of DM. Besides, the number of neutrinos ($\nu_e$) and protons ($p$), are also slightly enhanced in the stellar matter, particularly towards the core of the star where pressure is higher. Conversely, particles such as $n,\; e$, and $\mu$ show reduced fractions in the presence of DM towards the core. This suggests that DM decreases the isospin asymmetry, which impacts the microscopic properties of the stellar matter. Given that the isospin asymmetry is expressed as $\delta = (n_n-n_p)/(n_n+n_p)$, where $n_n$ is the neutron number density and $n_p$ is the proton number density. Generally, the value of $\delta$ increases as the star evolves from the first to the fourth stage, in the direction of deleptonization. For conventional NSs composed primarily of nucleons, $\delta$ is close to unity because neutrons dominate its composition. Additionally, the nuclear symmetry energy, a critical component of the EoS, is proportional to $\delta^2$ and quantifies the energy cost associated with the imbalance between proton and neutron numbers.

Furthermore, the particle abundances in the matter depend on the entropy density and lepton fraction, particularly the hyperonic species' appearance. In the first panel, we have a higher lepton fraction and lower $s_B$, therefore, the temperature of the stellar matter is lower than the intermediate stages (second and third panels) during neutrino diffusion as we shall see later (in Fig.~\ref{fig:RxT}). Here, the $\Lambda$-hyperons remain prominent in the stellar matter as in the other snapshots, but their appearance is relatively delayed to $R\sim 0.86R_{2.1}$,  toward the core, with $5\%$ DM mass fraction compared to the other snapshots.
The other hyperonic species ($\Xi$ and $\Sigma^-$) are less prominent and only appear deeper in the core from $R\sim 0.66R_{2.1}$ inwards, with $\Sigma^+$ and $\Sigma^0$ not appearing in significant quantities. At $T=0$ (last panel), on the other hand, we have even fewer hyperon species in the matter, only $\Lambda$-hyperon, $\Xi^{0}$, and $\Xi^{-}$ appear in significant amounts and their appearance is further delayed than in the first stage. Here, the $\Lambda$-hyperon starts appearing at $R\sim 0.66R_{2.1}$ and the $\Xi$ species at $R\sim 0.45R_{2.1}$ for $5\%$ DM mass fraction and deeper towards the core as the DM content reduces, the $\Sigma$'s do not appear at this stage. At the intermediate stages, where $s_B$ increases and $Y_{L,e}$ decreases (leading to an increase in temperature), all hyperonic species emerge quickly near the star's surface. At $R\sim 0.74R_{2.1}$ all the hyperons had appeared in the matter. The most prominent hyperon, the $\Lambda$-hyperon appear at $R\sim 0.92R_{2.1}$ in the second stage and $R\sim 0.93R_{2.1}$ in the third stage. Consequently, the production of hyperonic species is thermally favored~\cite{Issifu:2023qyi, Raduta:2020fdn, Malfatti:2019tpg}. A detailed analysis of the particle distribution against density can be found in~\cite{Issifu:2023qyi} where the EoS was originally used to study the evolution of PNSs.

{It is important to mention that in the approach adopted here (and hereafter), the gravitational mass is fixed instead of the total baryon mass ($M_b$). The latter quantity represents the sum of dark and visible baryon masses for DANSs or only the visible baryon mass in the absence of DM and is expected to be conserved during PNS evolution \cite{Nakazato:2020ogl, Pons:1998mm}. As a result, the obtained stellar profiles do not necessarily correspond to the same star along the evolution lines. The corresponding values of $M_b$ have also been recorded in Tab.~\ref{SP}, and the results show that $M_b$ decreases as $F_D$ increases. This behavior has important physical implications. It suggests that DM contributes with an additional gravitational pull, allowing the NS to reach a stable configuration with fewer $M_b$. Since the baryonic matter (DM plus OM) becomes more compressed due to this gravitational pull, the $M_b$ required to reach a given gravitational mass decreases compared to the case of a lower $F_D$ or OM-only configuration. Additionally, since DM is assumed to be self-interacting, it contributes to the degenerate pressure within the NS. However, the overall compression remains significant, keeping the $M_b$ lower for stars with a higher $F_D$ relative to those composed entirely of OM or lower $F_D$. Consequently, a fewer $M_b$ is required to maintain hydrostatic equilibrium for a fixed gravitational mass (see Ref.~\cite{Kumar:2025ytm}, which discusses the impact of DM on the $M_b$).}

\begin{figure}[!t]
\centering
\includegraphics[width=\linewidth]{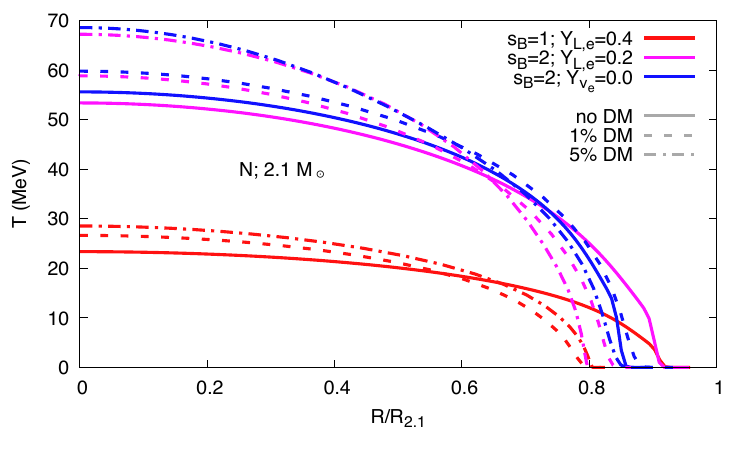}
\includegraphics[width=\linewidth]{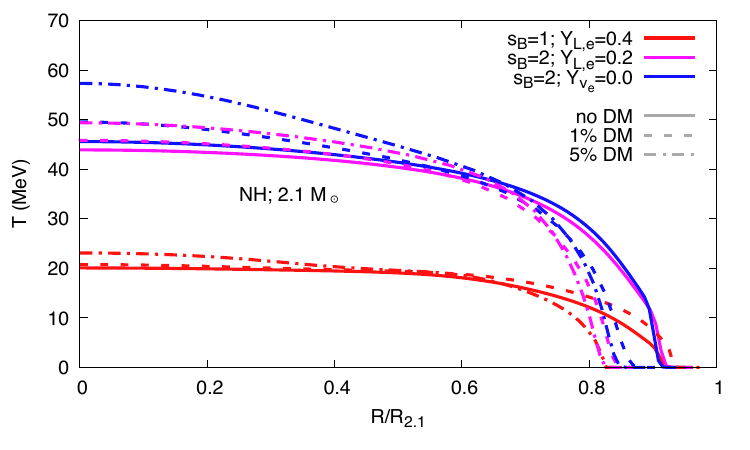}
\caption{Temperature profile of a $2.1M_\odot$ star, considering nucleonic (top) and hyperonic (bottom) hadronic matter compositions.
}
\label{fig:RxT}
\end{figure}

We present the effect of DM on the temperature distributions in the stellar matter in Fig.~\ref{fig:RxT}. 
Generally, the increase in the degree of freedom of the nuclear matter leads to a decrease in temperature, since we keep the entropy density fixed whilst increasing particle degrees of freedom. This accounts for the differences in the curves of the top panel (nucleon only) and the bottom panel (nucleon plus hyperons). Additionally, the presence of DM leads to an increase in the temperature of the stellar matter. This is because the gravitational potential of the NSs increases due to the presence of the DM (as shown in Fig.~\ref{fig:mxr}). The NSs with more DM are more compact so the stellar matter is more compressed. The compression increases the temperature of the matter due to the release of gravitational potential energy, in accordance with the virial theorem. According to this theorem, the kinetic energy (in this case, temperature) is related to the potential energy in a gravitationally bound system ($2T+U=0$, where $U$ is the gravitational potential energy)~\cite{terebizh2024evolution, CarrollOstlie2006}. The first stage of the PNSs evolution $(s_B=1;\; Y_{L,e}=0.4)$ is when the temperature is least increased by the presence of DM, suggesting that the first instants of the birth of the PNS do not present a big difference in temperature due to DM. On the other hand, during the second and third stages, as the star loses its neutrinos and the entropy increases $(s_B=2)$, the presence of DM causes a greater change in the temperature profiles. Thus, a few seconds after the formation of the PNS, the DM content begins to significantly influence temperature profiles by enhancing it, especially towards the core. It is interesting to note that in nucleonic PNSs, the increase in temperature from the second to the third stage of evolution is smaller when dark matter is present. In contrast, for hyperonic stars, a higher DM content results in a greater increase in temperature between these stages. This can be attributed to the compact nature of the star when hyperons are present, which becomes even more pronounced due to the presence of DM. 
Thus, the presence of DM alters the star's thermal equilibrium state, potentially prolonging its cooling phase during evolution. This effect could challenge our understanding of the star's age and history of thermal evolution.

\begin{figure}[!t]
\centering
\includegraphics[width=\linewidth]{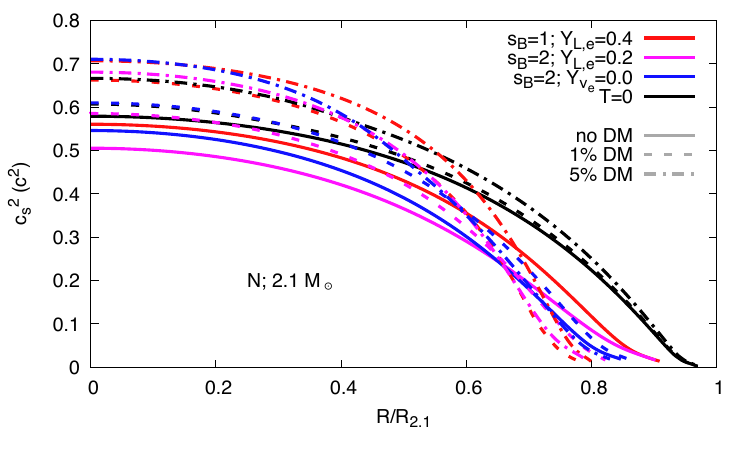}
\includegraphics[width=\linewidth]{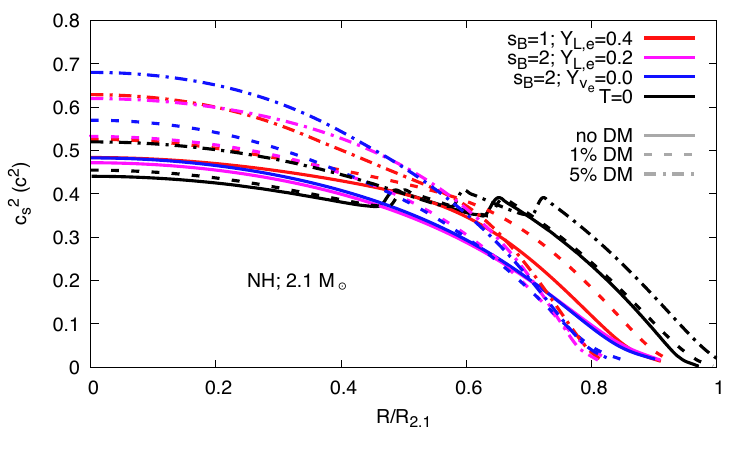}
\caption{Speed of sound squared profile of a $2.1 M_\odot$ star, considering nucleonic (top) and hyperonic (bottom) hadronic matter compositions admix DM.
}
\label{fig:Rxc}
\end{figure}

{Using the EoS for the OM, we can calculate the speed of sound ($c_s^{OM}$) directly from $(\sqrt{dP_{OM}/d\varepsilon_{OM}})|_{s_B}$. Thus, for each pressure value $P_{OM}$, we obtain the corresponding $c_s^{OM}$ for the OM. We then estimate the speed of sound $c_s$ (for the two fluids) at any given $P(r)$ in the DANSs within the known values of $c_s^{OM}$ using a cubic spline interpolation method for each point in $P_{OM}(r)$.  Additionally, considering that the DM mass fraction relative to OM is small (1\% and 5\%), this approach serves as a reasonable first approximation in the absence of a rigorous first-principles approach. Particularly, because the two-fluid TOV system changes the internal pressure balance which is accounted for in this approach. However, an alternative approach that determines an `effective speed of sound' by incorporating contributions from both matter components and studying the impact of DM on an `effective' $c_s$ tuned by a constant parameter that regulates the DM content can be found in Ref.~\cite{Giangrandi:2022wht}.}

In Fig.~\ref{fig:Rxc}, we calculate the square of sound speed ($c_s^2 = \partial P/\partial\varepsilon$) in the star's interior as a function of radius. 
This enables us to investigate the effect of the presence of DM on the EoS of the DANSs using a star of $2.1 M_\odot$ as a medium. Generally, softer EoSs lead to a decrease in $c_s^2$ relative to stiffer ones~\cite{Moustakidis_2017, Bedaque_2015}. Comparing, the top panel (nucleon only) and the bottom panel (nucleon plus hyperons) we observe that an increase in degree of freedom leads to a decrease in the $c_s^2$. This is consistent with softening the EoS if the degree of freedom of the stellar matter increases~\cite{Glendenning:1991es, Weissenborn:2011kb}. Comparing the curves in each panel for the different proportions of DM, we observe that DM increases the $c_s^2$. This implies that the presence of DM increases the core pressure of the DANSs. DM inside an NS is expected to increase its gravitational binding energy, which can compress the star and increase its core density. Therefore, in higher-density media, the presence of the DM increases the repulsion at short distances between the particles which could account for the rise in $c_s^2$. The presence of DM also affects the equilibrium composition of the star. This is evident in Figs.~\ref{fig:yxr} and~\ref{fig:RxT}. In Fig.~\ref{fig:yxr}, the presence of DM decreases the isospin asymmetry impacting the entire particle distribution in the star. In Fig.~\ref{fig:RxT}, the presence of DM increases the core temperature leading to changes in the thermal equilibrium state. These changes can significantly alter the EoS, leading to an increase in $c_s^2$. High $c_s^2$ is associated with the ability of the star to resist high gravitational pressures without collapse, which could impact the maximum mass, radius, and tidal deformability thresholds~\cite{Lattimer:2000nx, Altiparmak:2022bke}.

In the bottom panel, where hyperons are present in the stellar matter, we observe distinct bumps in the $c_s^2$ curves (black lines).
These bumps indicate shocks in the particle layers, occurring after the star has cooled down. For instance, we see two bumps in $c_s^2$ at $T=0$, which correspond to the appearance of two hyperonic species $\Lambda$ and $\Xi^-$ at $R \sim 0.62\rm R_{2.1}$ and $R \sim 0.38\rm R_{2.1}$ for DANSs with 5\% DM mass fractions respectively. This value shifts towards the core when the DM mass fraction is reduced, this is evident in Fig.~\ref{fig:yxr}. A clear understanding of the role of DM in changing the $c_s^2$ and the EoS of an NS could serve as a source for explaining the discrepancies in NS mass-radius measurements and help in the search for DM signatures in NSs.

\begin{figure}[!t]
\centering
\includegraphics[width=\linewidth]{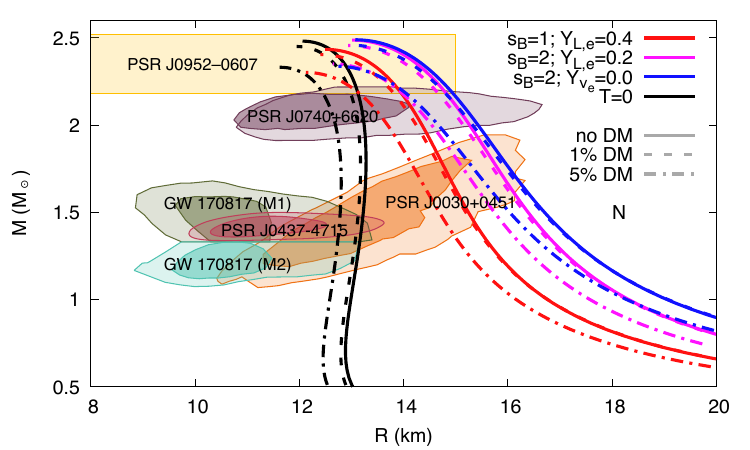}\\
\includegraphics[width=\linewidth]{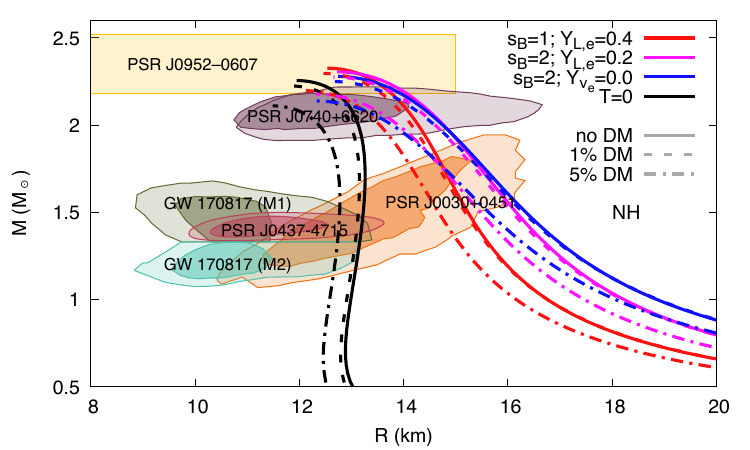}
\caption{Total gravitational mass ($M$) of a PNS as a function of the radius ($R$). The top panel illustrates the evolution of neutrino-trapped, $\beta$-equilibrated stellar matter at various stages of the star's evolution, characterized by different entropy per baryon ($s_B$) and lepton fraction ($Y_{L,e}$), compared to a neutrino-transparent star with $s_B = 2; \; Y_{\nu_e}=0$ and $T=0$ for nucleonic composition. The bottom panel shows the same scenario but with the inclusion of hyperonic degrees of freedom. Observational data: PSR J0740+6620~\cite{NANOGrav:2019jur, Fonseca:2021wxt}, PSR J0030+0451~\cite{riley2021}, PSR J0437-4715\cite{Choudhury:2024xbk}, PSR J0952-0607~\cite{Romani:2022jhd}, GW170817~\cite{GW170817_MR_PEsample, LIGOScientific:2017vwq}.}
\label{fig:mxr}
\end{figure}

In Fig.~\ref{fig:mxr} we show the gravitational mass as a function of the radius for DM-admixed PNSs. 
The top panel depicts the evolution of $\beta$-equilibrated PNSs composed of nucleons at different evolutionary stages, defined by different $s_B$ and $Y_{L,e}$. The bottom panel shows stars with hyperons in their core at different evolutionary stages similar to the nucleonic ones. In the absence of DM, represented by the solid lines in both panels, the stellar configuration is determined solely by the nucleonic or hyperonic EoSs. From the figure, we see that the maximum mass and its respective radius, both related to the stage in which $Y_{L,e}=0.4$, are given by $M_{\rm max}=2.43 M_{\odot}$ 
and $R_{\rm max}=12.49$~km. Furthermore, the radius of the canonical star is $R_{1.4}=15.41$~km. With increasing entropy and the onset of deleptonization, the stars get heated and expand. Therefore, $R_{1.4}$ and $R_{\rm max}$ increase without a significant change in $M_{\rm max}$, as displayed in Tab.~\ref{SP}. 


At $T=0$ the stars shrink, so their radii decrease. When DM is introduced, represented by dashed and dot-dashed lines for 1\% and 5\% respectively, an increase in DM mass fraction leads to a decrease in the maximum mass and radius of NSs. The same phenomenon is observed in all stages of the star's evolution. The presence of DM enhances the gravitational interaction at the star's center. Consequently, mass is drawn inward, leading to increased central baryonic density and reduced star's radius. This results in the star's compactification. The degree of compactification due to DM presence is approximately the same for all four stages of evolution considered, so the temperature of the PNSs does not significantly affect how much the star shrinks due to DM.

The mirrored DM does not contribute significantly to the degeneracy pressure that supports the star from gravitational collapse due to its nature of interaction with the OM. This reduces the star's capacity to resist compression, thereby lowering its mass and radius simultaneously~\cite{de2010neutron, Goldman:1989nd}. Core clustering of NSs due to DM accumulation also displaces the OM which is meant to provide degeneracy pressure to support the star from collapse leading to a decrease in mass and radius. The presence of DM alters the thermal behavior of the DANSs (as discussed in Fig.~\ref{fig:RxT}) by contributing to heat dissipation. This has an effect on the size of the star during its evolutionary stages. 
From the particle population in Fig.~\ref{fig:yxr}, we observe that the presence of DM reduces $\delta$. 
These combined effects produce relatively more compact DANSs with lower mass and radius in the presence of fixed DM mass fraction. 
The observed deviation of the mass-radius relationship from the predicted values based on pure ordinary matter 
can serve as a means to indirectly probe the presence of mirror dark matter in neutron stars. However, it is important to mention that the results presented here are model-dependent.

\begin{figure}[!t]
\centering
\includegraphics[width=\linewidth]{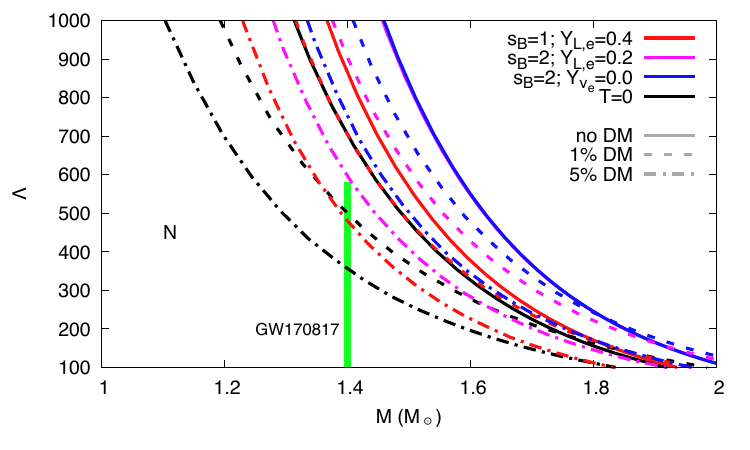}\\
\includegraphics[width=\linewidth]{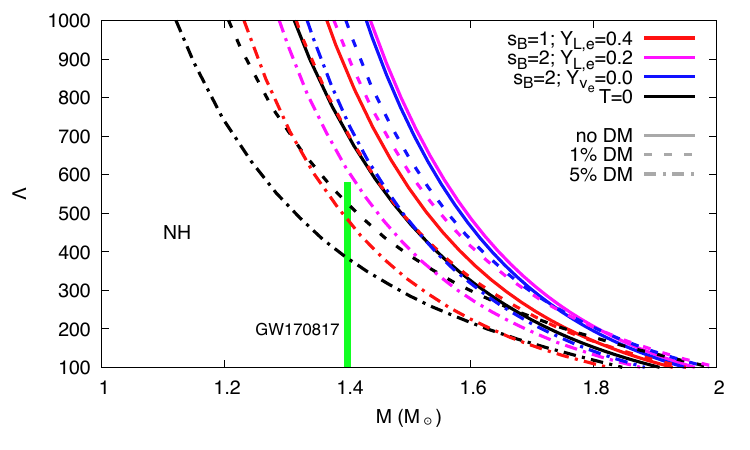}
\caption{Dimensionless tidal deformability ($\Lambda$) as a function of the total gravitational mass ($M$) in solar masses ($\rm M_{\odot}$). The  vertical {green} bar represents the tidal deformability constraints ($\Lambda_{1.4} = 190^{+390}_{-120}$
) at $\rm 1.40 M_{\odot}$ from the GW170817 observation~\cite{LIGOScientific:2018cki}. The top panel and bottom panels are nucleonic and hyperonic matter respectively.}
\label{fig:mxl}
\end{figure}

In Fig.~\ref{fig:mxl}, we present the results for the dimensionless tidal deformability parameter as a function of the stellar mass for the PNSs admixed DM from birth to ``maturity''. It measures how easily the shape of the star is deformed under the influence of an external gravitational field. Similar to the other figures discussed above, the top panel represents nucleonic matter and the bottom panel represents hyperonic matter, both cases are mixed with $1\%$ and $5\%$ fixed DM mass fractions. The figures in both panels are similar because of the density-dependent emergence of hyperons. These particles begin to appear only in the densest regions near the core of the star. Since the core represents a relatively small portion of the total stellar volume, and tidal deformability is more sensitive to the outer layers of the star, the presence of hyperons has a negligible effect on the star's ability to resist external tidal forces (see~\cite{Kumar:2016dks} and references therein for further discussions). The focus of showing the tidal deformability value measured for the GW170817 event is for comparison with the cold EoS. However, the results for the fixed entropy stars can serve as projections for future gravitational wave observations, since new detectors such as Einstein Telescope~\cite{ET:2019dnz} and Cosmic Explorer~\cite{LIGOScientific:2016wof}, which are currently in development, will be able to detect signals during the post-merger phase. 

From the top and bottom panels, we observe that the DM mass fractions reduce $\Lambda$~\cite{Das:2021wku}, the same way they reduce gravitational mass and radius. This is attributed to changes in the internal (see Fig.~\ref{fig:yxr}) and external structures (see Fig.~\ref{fig:mxr}) of the star due to DM, thus changing its response to external tidal forces~\cite{Leung:2022wcf, Koehn:2024gal}.  The $\Lambda$ depends inversely on the compactness of the star,  which is the ratio of the gravitational mass to the radius, as shown in Eq.~(\ref{td}). 
Therefore, the decrease in $\Lambda$ with increasing DM mass fraction indicates that the presence of DM enhances the star's compactness, as the mass and radius data in Tab.~\ref{SP} shows. Aside from the 1\% and 5\% cold-catalyzed DANSs satisfying the tidal deformability value of GW170817, we observe that a newly born PNS ($s_B =1; Y_{L,e}=0.4$), with relatively low temperature, satisfies the GW170817 constraint for $5\%$ DM mass fraction. 
However, all the fixed-entropy stars without dark matter content strongly violate the $\Lambda$ value demarcated for the GW170817 event.
As a result, hotter stars are more easily deformed compared to their relatively cooler counterparts. This is attributed to the thermal pressure generated as a result of the increase in temperature, leading to an increase in energy density and particle motion in the stellar matter. This compromises the structural rigidity of the star, making it more prone to external forces~\cite{Wei:2021veo, Lu:2019mza}.

\section{Final Remarks}\label{fm}
This work investigates how the presence of DM in the core of a PNS alters its microscopic and macroscopic properties, from its birth as a neutrino-rich object to its evolution into a cold, catalyzed, neutrino-poor star. We consider a simple mirror DM model, where $1\%$ and $5\%$ of DM mass fraction accumulate in the stellar core. The amount of dark matter present within a neutron star is largely determined by the rate at which dark matter particles are captured. This capture rate, influenced by the nature of interactions between dark matter and hadrons in the star's dense core, can reach values around $10^{25} \, \text{GeV/s}$ for dark matter particles with a mass near 1 GeV~\cite{Bell:2020jou}. Considering the neutron star’s lifetime of approximately $10^{17}$ seconds, such rates imply that the accumulation of dark matter within neutron stars is likely insufficient to form a significant fraction of their total mass. This informed the choice of the DM mass fractions used in this work.

We determined that the presence of DM influences the particle distributions, leading to a decrease in the isospin asymmetry and the early emergence of hyperonic particles. 
The presence of DM enhances the population of neutrinos, protons, and hyperonic species towards the stellar core while neutrons, electrons, and muons are reduced in the same direction. The effect of DM on the particle distribution is more pronounced in relatively colder stars (see first and last stages of Fig.~\ref{fig:yxr}) where fewer hyperons appear compared to the hotter ones. These changes in the internal structure of the PNSs are reflected in the thermal profiles, the sound speed, the mass-radius relation, and the tidal deformability of the star. We combined advanced EoS modeling with observational constraints to demonstrate how DM fundamentally alters both the internal structure and observable features of compact objects using a two-fluid approach. 

The main findings are summarized below:
\begin{itemize}
  
    \item In Fig.~\ref{fig:RxT}, the results show that the presence of DM heats up the star. The increase in temperature is linked to the compression caused by DM-induced gravitational potential energy, which modifies the thermal history and potentially prolongs the cooling phase of the star. 
    The presence of dark leptons also affects the cooling dynamics of the star, as discussed in~\cite{Reddy:2021rln}. Aside from that, the presence of DM in hyperonic stars leads to a higher jump in temperature in the intermediate stages (second and third stages) when the star is relatively hotter. These alterations in thermal equilibrium due to DM presence challenge the conventional assumptions about the age and evolutionary trajectory of NSs.
    
    \item In Fig.~\ref{fig:Rxc}, the $c_s^2$ is more sensitive to the increase in core pressure due to the compactness caused by the presence of DM. On the other hand, the mass-radius relation is less sensitive because the gravitational pull from the DM pulls mass inward causing a reduction in radius. Additionally, since the presence of the DM does not contribute to degenerate pressure that supports the star against collapse, it reduces the total baryonic mass the star can support before collapse, this lowers the maximum gravitational mass~\cite{Collier:2022cpr, Emma:2022xjs} as can be seen in Fig.~\ref{fig:mxr}. Thus, the compactness does not lead to stiffer EoS relative to the baryonic one. Notably, the formation of hyperonic species introduces distinct bumps in the $c_s^2$ profiles, indicating localized effects on the EoS~\cite{Issifu:2023ovi, Lopes:2023gzj}.

      \item The EoS satisfies the mass-radius measurement of the pulsars: PSR J0740+6620~\cite{NANOGrav:2019jur, Fonseca:2021wxt} and PSR J0030+0451~\cite{riley2021} with their respective contours demarcated on Fig.~\ref{fig:mxr}. Both fixed entropy stars and the cold stars with or without DM satisfied PSR J0952-0607~\cite{Romani:2022jhd} mass constraint, except for the cold star with 5\% DM mass fraction and hyperons in its core. Other constraints satisfied are: PSR J0437-4715~\cite{Choudhury:2024xbk} and GW170817~\cite{GW170817_MR_PEsample, LIGOScientific:2017vwq} when the star is cold and catalyzed. 
  
\end{itemize}

In conclusion, changes in the internal structures of PNSs are evident in particle distributions, thermal fluctuations, and sound speed variations within the stellar matter. These changes manifest as observable discrepancies in the physical properties, such as mass, radius, and tidal deformability. {These insights pave the way for future studies, aiming to indirectly detect DM through astrophysical observations, offering a promising intersection of nuclear astrophysics and particle physics. }

\begin{acknowledgments}
P.T. sincerely thanks Davood Rafiei Karkevandi for his useful discussions on dark matter. This work is part of the project INCT-FNA proc. No. 464898/2014-5. A.I. acknowledges financial support from the São Paulo State Research Foundation (FAPESP), Grant No. 2023/09545-1. T. F. thanks the financial support from the Brazilian Institutions: CNPq (Grant No. 306834/2022-7), Improvement of Higher Education Personnel CAPES (Finance Code 001), and FAPESP (Grants No. 2017/05660-0 and 2019/07767-1). 
D.P.M. is partially supported by CNPq (Grant No. 303490/2021-7).
K.D.M. is supported by FAPESP, Grant No. 2024/01623-6.  O.L. and M.D. are supported by CNPq under Grants No. 307255/2023-9, and No. 308528/2021-2, respectively, and the project No. 01565/2023-8~(Universal). Kau Dalfovo Marques died just two days after the acceptance of this paper, at the age of 30.  Despite her neurological disease, she fought for her life and science with excellent humor. She will be missed by everyone fortunate to work with her and will certainly remain an eternal inspiration to all of us.
\end{acknowledgments}

\bibliography{references}
\end{document}